\documentclass{aa}

\usepackage{graphicx}
\usepackage{soul}
\usepackage{amsmath}
\usepackage{txfonts}
\def\aL13{A_\textrm{L13}}
\def\C0{C^0}
\def\Cref{C^{\mathrm{ref}}}
\newcommand{\bigdot}[1]{\overset{\mbox{\boldmath $.$}}{#1}}

\usepackage{nidanfloat} 

\begin{document}

\title{The amplitude of the cross-covariance function of solar oscillations as a diagnostic tool for wave attenuation and geometrical spreading}

\titlerunning{The amplitude of the cross-covariance function}

\author{Kaori Nagashima\inst{1} \and 
Damien Fournier\inst{2}
 \and  Aaron C. Birch\inst{1}
 \and Laurent Gizon\inst{1,3}}
\institute{
Max-Planck-Institut f\"ur Sonnensystemforschung, 
 Justus-von-Liebig-Weg 3, 37077 G\"ottingen, Germany\\
\email{nagashima@mps.mpg.de}
 \and
  Institut f\"ur Numerische und Angewandte Mathematik, Georg-August-Universit\"at G\"ottingen, Lotzestra{\ss}e 16-18, 
  37083 G\"ottingen, Germany
 \and
  Institut f\"ur Astrophysik, Georg-August-Universit\"at G\"ottingen, 
  Friedrich-Hund-Platz 1, 37077 G\"ottingen, Germany  
}

   \date{Received October 5, 2016; accepted December 2016}

\abstract
   {In time-distance helioseismology, wave travel times are measured from the two-point cross-covariance function of solar oscillations and are used to image the solar convection zone in three dimensions. There is, however, also information in the amplitude of the cross-covariance function, for example about seismic wave attenuation.}
   {Here we develop a convenient procedure to measure the amplitude of the cross-covariance function of solar oscillations.}
   {In this procedure, the amplitude of the cross-covariance function is linearly related to the cross-covariance function and can be measured even for high levels of noise.}
   {As an example application, we measure the amplitude perturbations of the seismic waves that propagate through the sunspot in active region NOAA~9787. We can recover the amplitude variations due to the scattering and attenuation of the waves by the sunspot and associated finite-wavelength effects.}
 {The proposed definition of cross-covariance amplitude is robust to noise, can be used to relate measured amplitudes to 3D perturbations  in the solar interior under the Born approximation, and will provide independent information from the travel times.}


\maketitle

\section{Introduction}\label{sec:intro}

Solar oscillations are excited stochastically by turbulent convection and can be used to probe solar interior dynamics and structure \citep[e.g.,][]{JCDLecNot2003, 2005LRSP....2....6G, 2010ARA&A..48..289G}. 
The starting point in time-distance helioseismology is the temporal cross-covariance function of the observed oscillation signal:
\begin{equation}
C(\vec{x}_1, \vec{x}_2,t) = \frac{1}{T}
\int^{T/2}_{-T/2}  \phi(\vec{x}_1,t^\prime) \phi(\vec{x}_2,t^\prime+t) \mathrm{d}t^\prime \ , 
\label{eq:CCdef}
\end{equation}
where the $\phi(\vec{x}_i,t^\prime)$ are the oscillation signals (e.g., Doppler velocity observations)  at time $t^\prime$ and positions $\vec{x}_i$ ($i=1,2$) on the solar surface, $t$ is the time lag,
and $T$ is the duration of the observation
\citep{1993Natur.362..430D,1997SoPh..170...63D}. Throughout 
this paper, for the sake of simplicity, we will treat the observable $\phi$ as a  continuous function of time. 

From the cross-covariance, one then generally measures wave travel times \citep[e.g.,][]{1997SoPh..170...63D, 2002ApJ...571..966G,2004ApJ...614..472G, 2015A&A...581A..67L} 
in order to constrain physical properties in the solar interior. The cross-covariance function, however, contains additional information other than the travel times 
\citep[e.g.,][]{2003ApJ...593.1242T}. 
It is possible to use all the information in the cross-covariance by performing a full waveform inversion; this approach is widely used in geophysics \citep[see e.g., the review of][]{virieux2009overview} and has been attempted in helioseismology by, for example,  
\cite{2014ApJ...784...69H}, but only in rather simple cases. Here,  instead of full waveform inversion, in addition to the travel time, we extract a second parameter from the cross-covariance: the amplitude.

\subsection{Previous measurements of amplitudes}

In the field of helioseismology, only a few studies have reported measurements of the amplitude of the cross-covariance function. 
\citet{2006ApJ...648L..75J} used Michelson Doppler Imager \citep[MDI;][]{1995SoPh..162..129S} observations to show that the amplitude of the cross-covariance is reduced for waves propagating away from an active region. In this study, amplitudes were measured by fitting the measured cross-covariance function with a Gabor wavelet \citep{1997ASSL..225..241K}.
Similar reduction of the cross-covariance amplitude is 
shown for an example plage region in 
\citet{2009ASPC..415..417N} 
but without a quantitative calculation of amplitudes.
\citet{2007SoPh..241...17B} measured the amplitude of the cross-covariance function (Gabor-wavelet fitting), and used the exponential decrease of the amplitude with the number of skips to estimate the solar p-mode lifetime.

\citet{2008SoPh..251..291C} compared forward-modeled and observed f-mode cross-covariances for an almost circular sunspot in active region NOAA 9787.  For a parametric magnetohydrostatic sunspot model with an axisymmetric magnetic field,
they found good agreement in both the amplitude and phase of the forward modeled and observed cross-covariances when the magnetic field peak at the photospheric level is 3kG  \citep[see also][]{2010ARA&A..48..289G}. 
We will use observations of this region in the current study (see Sect.~\ref{sec:example}).

\citet{2013A&A...558A.129L} measured the amplitude reduction of waves propagating through the same sunspot and used a 2D ray theory calculation to show that the amplitude reduction is, in part, due to the defocusing of waves caused by the enhanced wave speed in the sunspot.

\subsection{Motivation for a new definition}

Two definitions for the amplitude of the cross-covariance are used in helioseismology: 1) the amplitude parameter obtained from fitting a Gabor wavelet to the cross-covariance \citep[e.g.,][]
{1997ASSL..225..241K,2006ApJ...648L..75J,2007SoPh..241...17B} and 2) the maximum of the analytic signal associated with the cross-covariance between the observed cross-covariance and a reference cross-covariance 
\citep[][see our Sect.~\ref{sec:L13def} for details]{2013A&A...558A.129L}.  

These definitions link the amplitude to the cross-covariance via complex transformations.  It is thus complicated to compute the  sensitivity of the amplitude measurement to small changes in a model of the solar interior.  These sensitivity calculations (kernel calculations) are necessary for inversions that would use measured amplitudes to infer  subsurface physical conditions. 

\subsection{What could we learn from the amplitude?}

We expect that the amplitude of the cross-covariance function will be useful for investigating wave properties in the Sun. For example, the amplitude is a more natural physical quantity for measuring attenuation than travel times \citep[e.g.,][]{1997ApJ...485..890W, 2003ApJ...596L.263T}.

In Earth seismology, the amplitude of the seismogram signals (i.e., seismic waves from a single source) as well as the amplitude of the cross-covariance of seismogram signals (i.e., due to ambient seismic noise field) are used to infer physical properties in the interior of the Earth.

Amplitudes have been used to characterize the seismic signals caused by earthquakes \cite[e.g.][]{2005SeismicEarth_37,DAL06}.
\cite{2002GeoJI.150..440D} showed body-wave speed 
Fr\'echet kernels for travel times and amplitudes
of seismogram signals and a 1D reference Earth model.  The kernels for travel times have a ``banana-doughnut'' shape: travel times are insensitive to changes in the wave speed along the ray path but are sensitive to changes in the surroundings.  In contrast,  the amplitudes are most sensitive to the wave speed along the ray path.
\cite{2005SeismicEarth_37} computed kernels for a volcano (Mount Vesuvius) model as well: travel times and amplitudes provide independent information about subsurface structure.

The amplitude of the cross-covariance of seismogram signals (ambient seismic noise field), as in time-distance helioseismology, has also been used in Earth seismology. The ambient noise field is produced by, for example, ocean swells \citep{GGGE:GGGE1208} or interaction of storms with the ocean \citep{Gerstoft869}.  \cite{PRI11} showed, using several observation examples, that the amplitude of the cross-covariance is useful for detecting attenuation and temperature perturbations in  the interior.  \cite{LIU13} modeled the effect of attenuation on amplitudes and the travel times (phases in their terminology) of the cross-covariance function.

The amplitude of earthquake seismogram signals has been used to create attenuation maps of Earth interior \citep{ROM95, DAL08}. Amplitude measurements have also been used for highly nonlinear problems to infer the attenuation (more precisely, the quality factor $Q$) inside the volcanoes Mount St. Helens and Mount Vesuvius from the seismogram data \citep{2014JVGR..277...22D}. 

As another example, the amplitude of the cross-covariance of ambient seismic noise is discussed in \cite{GGGE:GGGE1208}. They used the signal-to-noise ratio of the cross-covariance to detect ambient noise sources.

We note that in one ambient seismic noise analyses,
\cite{2010GeoJI.181..985M} reported the detection of 
seasonal wave velocity variations in Los Angeles Basin using the travel-time shift measured from the cross-covariance function of the ambient seismic noise. 
Later \cite{2013GeoJI.194.1574Z} and
\cite{2016GeoJI.205.1926D} suggested that the observed effect is caused by seasonal variations in the frequency content of the wave sources and does not actually imply velocity variations.   This is another example where using amplitude information might help to improve understanding of a physical system.

In order to interpret amplitude measurements, a relationship between measured amplitudes and physical parameters such as attenuation or wave velocities is needed. As the first step, we propose a new definition of the amplitude of the cross-covariance function:
the formulation procedure is described in Sect.~\ref{sec:Formulation}, and the definition is given in Sect.~\ref{sec:Def}.
In this definition,
changes in the amplitude are linearly related to changes in the cross-covariance.  This linearity will greatly simplify the calculation of sensitivity kernels and thus open the possibility of using amplitudes as inputs for inversions. It also has advantages in terms of the noise behavior as described in 
Sect.~\ref{sec:CompL13}. To show that our proposed definition is reasonable, in Sect.~\ref{sec:example} we compare the amplitude measurements of \cite{2013A&A...558A.129L} 
with the amplitudes obtained from our definition. Conclusions and outlooks of this study are given in Sect.~\ref{sec:conclusions}.

\section{Measuring the amplitude of a noise-free cross-covariance function}
\label{sec:Formulation}

\begin{figure*}
\centering
\begin{tabular}{cc}
\includegraphics[height=6cm]{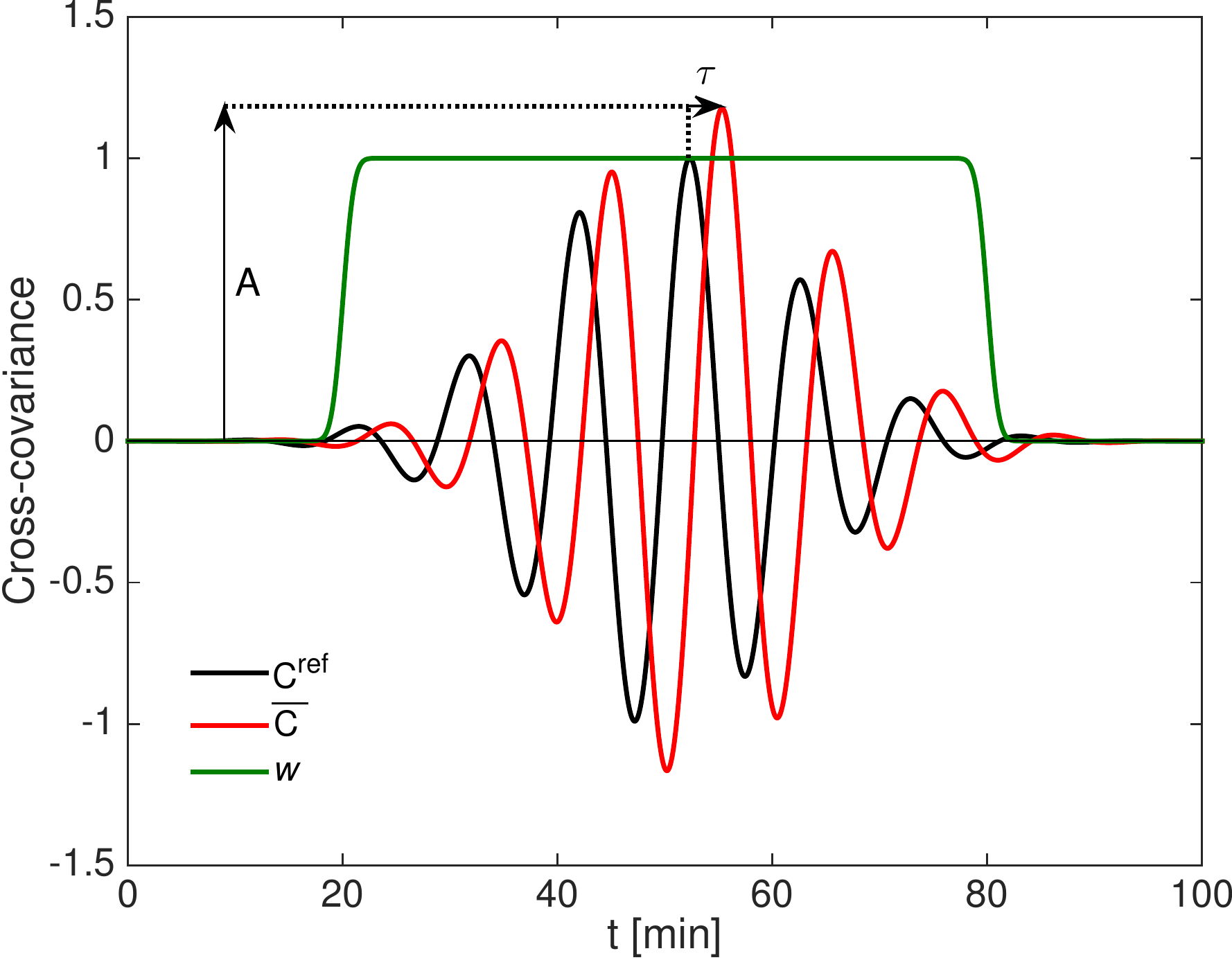} & \includegraphics[height=6cm]{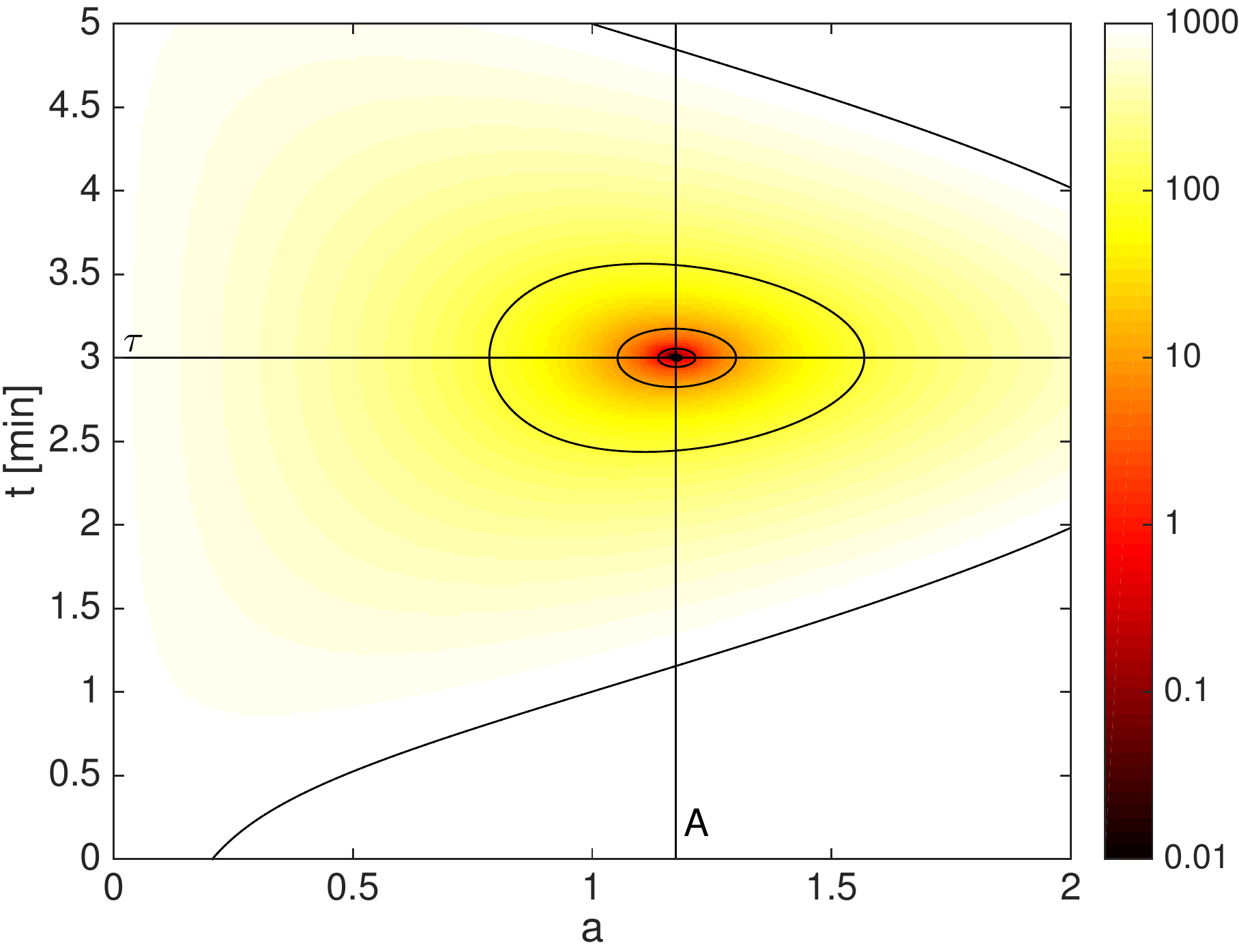}
\end{tabular}
\caption{Left: Illustrative sketch of a perturbed and reference cross-covariances ($\overline{C}$ and $\Cref$, respectively) and the window function $w$.
Right: The associated cost function $X(a,t)$ in log-scale color with contours where $X= 0.1$, $1$, $10$, $100$ and $1000$.
The color scale gives the value of the misfit.
To measure $\tau$ and $A$, we need to minimize $X$,
and in this case the minimum is reached for $\tau = 3$~min and $A = 1.2$. 
We note that in this plot (left panel) we show only $t>0$.
This part of the cross-covariance function corresponds to waves propagating  
from the position $\vec{x}_1$ to $\vec{x}_2$, and the window function for this wave is zero for $t<0$.}
\label{fig:toyPlot}
\end{figure*}

\cite{2002ApJ...571..966G} define the travel time $t=\tau$ 
as the time lag which minimizes the cost function
\begin{eqnarray}
X(t) = \int^{+\infty}_{-\infty} w(t^\prime) \left[\overline{C}(t^\prime) - \Cref(t^\prime-t) \right]^2  \mathrm{d}t^\prime \ , \label{eq:GB1para_chisq} 
\end{eqnarray}
where $w$ is a window function, $\overline{C}$ is a noise-free 
cross-covariance, and $\Cref$ is a reference cross-covariance. 
The noise-free cross-covariance $\overline{C}$ can be understood as the expectation value of $C$ (see Sect.~\ref{sec:Def}).
In Eq.~(\ref{eq:GB1para_chisq}) and hereafter, we drop $\vec{x}_1$ and $\vec{x}_2$ from $C(\vec{x}_1, \vec{x}_2,t)$ 
and write $C(t)$ to simplify the notation.
It is reasonable to use the noise-free cross-covariance
when the observation duration, $T$, is long enough.  The window function $w$ is chosen to select the target wavepacket.  As an example, one choice could be $w(t)=1$ at the time lags $t$ where the target wavepacket has significant amplitude and a smooth transition to zero outside this range of time lags (for an example, see Fig.~\ref{fig:toyPlot}). 

Here, we extend this one-parameter fit to a two-parameter fit for an amplitude, $a=A$, and a travel time, $t=\tau$ which minimize the cost function 
\begin{eqnarray}
X(a,t) = \int^{+\infty}_{-\infty} w(t^\prime) \left[\overline{C}(t^\prime) - a \Cref(t^\prime-t) \right]^2 \mathrm{d}t^\prime  \ . \label{eq:2para_chisq}
\end{eqnarray}

Figure~\ref{fig:toyPlot} shows example cross-covariances $\overline{C}$ and $\Cref$ and the associated cost function $X(a,t)$. In this Figure, $\overline{C}(t) = 1.2 \Cref(t-3~\rm{min})$  and the minimum of $X$, $X(A,\tau)=0$, is reached for $\tau = 3$~min and $A = 1.2$ 
as expected. 

As described in the introduction (Sect.~\ref{sec:intro}), we would like to have a definition in which changes in the amplitude are linearly related to changes in the cross-covariance.
Thus, we linearize Eq.~(\ref{eq:2para_chisq}) using
\begin{eqnarray}
\overline{C}(t) = \C0(t)+\delta \overline{C}(t) \ , \ a =a^0 + \delta a  \  , \  t = t^0 +\delta t
\end{eqnarray} 
and assume the perturbations (with $\delta$) are small. This is justified when the reference cross-covariance is close enough to the observed one, that is to say, the perturbation to the solar model is small.  
For the sake of simplicity, we will consider the case where
\begin{equation}
\Cref=\C0 
\end{equation}
and is even in time. In this case, the zeroth-order terms of Eq.~(\ref{eq:2para_chisq}) give $a^0=A^0=1$ and $t^0=\tau^0  = 0$.  Appendix~\ref{sec:ap_2parader} shows the details of the calculation.
The cross-covariance $\C0$ is obtained either from a model or from a spatial average of the observed cross-covariance function.

If we choose a window function $w(t)$ to select a particular wavepacket (for example, a smooth box like in Fig.~\ref{fig:toyPlot}),
the minimum of Eq.~(\ref{eq:2para_chisq}), at  first order, occurs at 
$a=A^0+\delta A= 1+ \delta A$ and $t=\tau^0+\delta \tau = 0 + \delta \tau$, where
\begin{eqnarray}
\delta A &=& \int^{+\infty}_{-\infty} W_{A}(t)\,  \delta \overline{C}(t) \, \mathrm{d}t , \label{eq:F2para_da_WC}\\
\delta \tau &=& \int^{+\infty}_{-\infty} W_{\tau}(t) \, \delta \overline{C}(t) \, \mathrm{d}t  , \label{eq:F2para_dtau_WC}
\end{eqnarray} 
with the weight functions 
\begin{equation}
W_{A}(t) = \frac{w(t) \C0(t) }{\int^{+\infty}_{-\infty}   w(t^\prime ) [\C0 (t^\prime )]^2 \mathrm{d}t^\prime}      
 \label{eq:F2para_a_weight_new} 
 \end{equation}
 and
 \begin{equation}
W_{\tau} (t) = - \frac{w(t) \bigdot{C}{}^0(t) }{\int^{+\infty}_{-\infty}  w(t^\prime) [\bigdot{C}{}^0(t^\prime)]^2 \mathrm{d}t^\prime} 
\label{eq:F2para_t_weight_new} ,
\end{equation}
where $\bigdot{C}{}^0(t)$ denotes the time derivative of $\C0 (t)$. 

Thus, if the window function selects a wave packet properly, 
the measurements for travel-time and amplitude are decoupled.
In particular, Appendix~\ref{sec:ap_2parader} shows that
the two measurements ($\delta A$ and $\delta \tau$) are decoupled
when the integral $I_2$ (defined by Eq.~(\ref{eq:2para_I2})) in Eq.~(\ref{eq:2para_matrix}) is negligible. In this case, the weight functions $W_{A}$ and $W_{\tau}$ are identical to what we would obtain from two independent single-parameter fits. In particular, $W_\tau$ is identical to the definition obtained in \cite{2002ApJ...571..966G} by minimizing Eq.~(\ref{eq:GB1para_chisq}).

\begin{figure}
\centering
\includegraphics[width=0.9\linewidth]{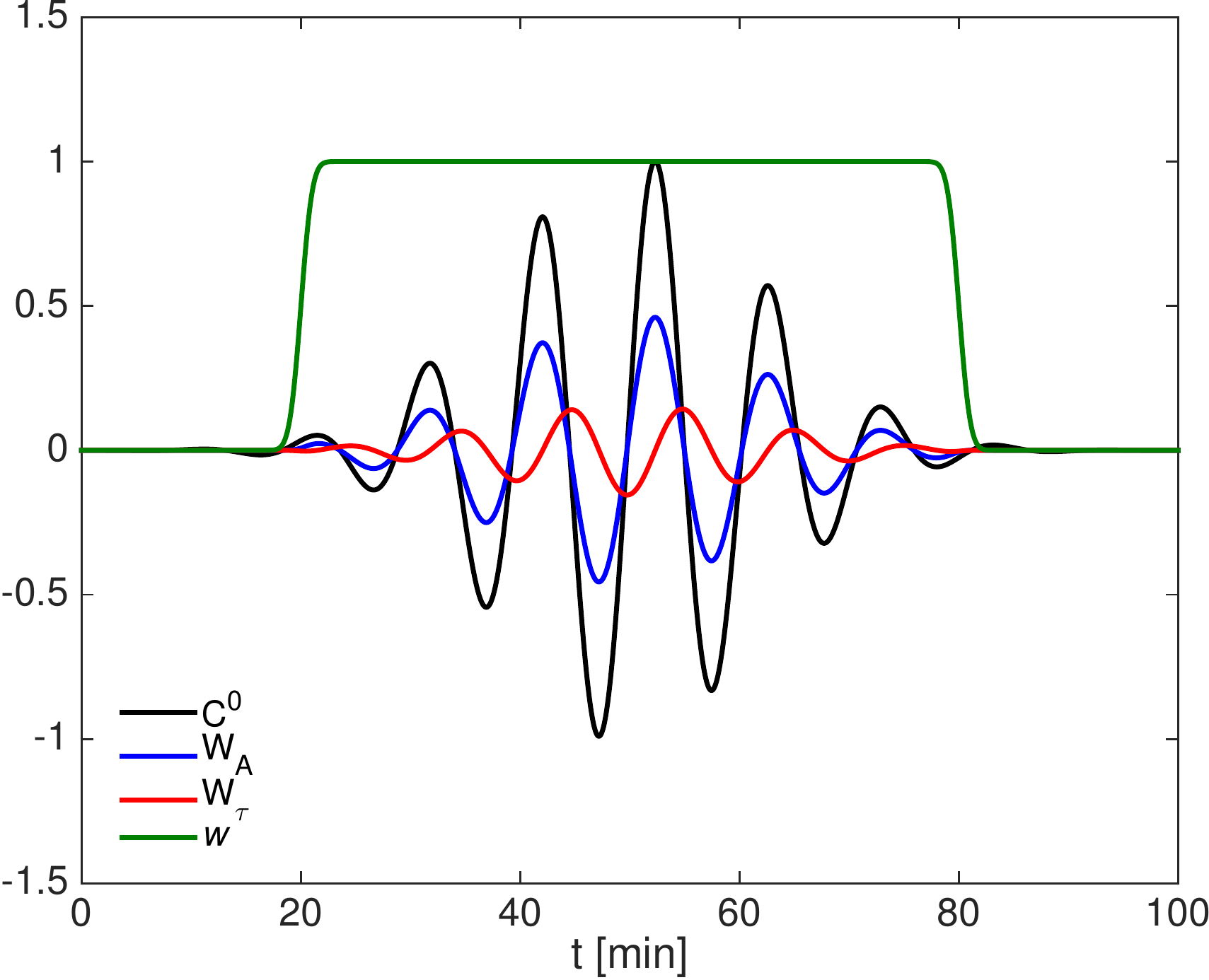}
\caption{
Illustrative sketch of a reference cross-covariance $\C0$ and the corresponding weight functions for the amplitude, $W_A$, and for the travel time, $W_\tau$,
as well as the window function, $w$. The scaling is arbitrary as the different quantities have different units.}
\label{fig:toyPlot_W}
\end{figure}

Figure~\ref{fig:toyPlot_W} illustrates the weight functions for amplitude and travel-time given by Eqs.~(\ref{eq:F2para_a_weight_new}) and (\ref{eq:F2para_t_weight_new}). Multiplying these weight functions by the change in the cross-covariance $\delta \overline{ C}$ and integrating over time lag  
give the travel-time and amplitude perturbations $\delta \tau$ and $\delta A$. 
The weight function $W_A$ is proportional to $\C0$ while $W_\tau$ is proportional to the derivative of $\C0$. 
Thus $W_A$ is in phase with $\C0$ while $W_\tau$ is shifted by a quarter of a period.

\section{Definition of the amplitude}
\label{sec:Def}

So far, we have minimized the cost function by supposing that the cross-covariance function is noise-free. 
This is the limit of the infinite length of observation duration, and is not the case 
in observations with finite duration. If the data are too noisy then the minimization of Eq.~(\ref{eq:2para_chisq}) would not make sense. In this case, similarly to \cite{2004ApJ...614..472G}, we can introduce a smoothed cross-covariance function $C^\epsilon$
\begin{equation}
C^\epsilon(t) = \epsilon C(t) + (1-\epsilon) \C0(t),
\end{equation}
where $C(t)$ is the observed cross-covariance function with noise
and $\epsilon$ is a small positive number, 
and perform the minimization of Eq.~(\ref{eq:2para_chisq}) with $C^\epsilon$ instead of $\overline{C}$. In Appendix \ref{sec:ap_GB04_2para} we show that in the limit of $\epsilon \rightarrow 0^+$, this procedure leads to the same expressions for the amplitude and the travel-time as given by Eqs.~(\ref{eq:F2para_da_WC}-\ref{eq:F2para_t_weight_new}).

Another way to see that the amplitude defined by Eq.~(\ref{eq:F2para_da_WC}) is applicable to a noisy cross-covariance is to use the linearity of the definition. 
Let us decompose the cross-covariance function with noise into the expectation value, $\overline{C}$, and the noise, $n(t)$,
\begin{equation}
C(t) = \overline{C} (t) + n(t) \; ,
\end{equation}
where the expectation value of the noise is zero: $\overline{n}(t)=0$.
Then, the expectation value of the amplitude is given by
\begin{eqnarray}
\overline{\delta A} = \int^{+\infty}_{-\infty}  W_A (t) \ \left( \overline{C}(t) -C^0(t) \right) \mathrm{d}t  \; .
\end{eqnarray}
Thus, the average value of the amplitude measured from the cross-covariance function with noise is the amplitude defined by the noise-free cross-covariance function.

Therefore, the expectation value of the amplitude 
of the observation cross-covariance function can be measured based on Eqs.~(\ref{eq:F2para_da_WC}) and (\ref{eq:F2para_a_weight_new}) as follows:
First, the cross-covariance $C^0$ is obtained either from a model or from a spatial average of the observed cross-covariance function.
Second, by using the window function for the targeted wave packet $w$, the weight function $W_A$ is obtained by Eq.~(\ref{eq:F2para_a_weight_new}).
The amplitude is then defined with the weight function as 
\begin{equation}
\boxed{\delta A = \int^{+\infty}_{-\infty} W_{A}(t) \delta C(t) \mathrm{d}t ,} \label{eq:amp_def}
\end{equation}
where $\delta C(t) = C(t) -\C0 (t)$. 
Hereafter in this paper we use Eq.~(\ref{eq:amp_def}) as the definition of 
the amplitude.
We note that this is different from Eq.~(\ref{eq:F2para_da_WC})
in that $\delta C$ contains noise.

This definition is analogous to the amplitude definition
in the Earth seismology analysis given by Eq.~(26) in \cite{2005SeismicEarth_37}. 

\begin{figure*}
\centering
\includegraphics[]{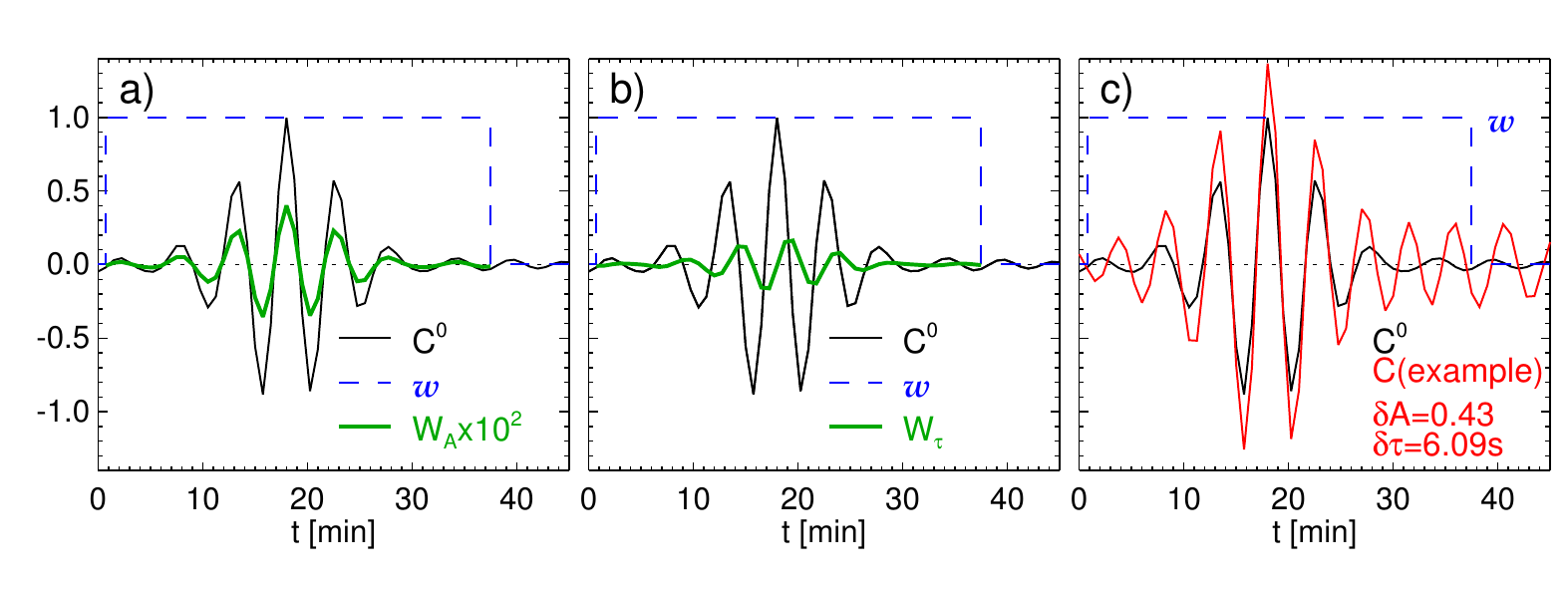}
\caption{p$_1$-mode cross-covariance functions (thin solid lines) and the weight functions (thick solid lines) calculated from a solar noise model datacube \citep{2015A&A...581A..67L} as well as the window functions (dashed lines), $w(t)$, to select the p$_1$-mode wave packet.
Panel a): the cross-covariance function averaged over the field of view as the reference cross-covariance function, $\C0$, 
and the weight function for amplitude $W_A$. 
Panel b): similar to a) but with the weight function for travel time $W_{\tau}$.
Panel c): one example cross-covariance function (at the central point of the field of view) as well as $\C0$.
The measured amplitude and the travel time for this example cross-covariance are written in the panel.  
The cross-covariance functions in these plots are normalized by the maximum of the absolute value of $\C0$.
The cross-covariance functions and $W_\tau$ are dimensionless;  the weight functions are calculated using the normalized cross-covariance and second as the unit of time. In the figure, $W_A$ is multiplied by $10^2$ for visibility, and the units are s$^{-1}$.}
\label{fig:cc_example_p1_noise}
\end{figure*}

\section{Comparison with the definition of \cite{2013A&A...558A.129L}}
\label{sec:CompL13}

\subsection{\cite{2013A&A...558A.129L}'s Definition} \label{sec:L13def}

Another way to measure the amplitude was proposed by \cite{2013A&A...558A.129L} in the context of sunspot seismology. The amplitude was obtained from the function
\begin{equation}
F(t) = \frac{\int^{+\infty}_{-\infty}  w(t^\prime) C(t^\prime) C^{0}(t^\prime-t)\mathrm{d}t^\prime}{\int^{+\infty}_{-\infty}  w(t^\prime) [C^{0}(t^\prime)]^2 \mathrm{d}t^\prime } \ , \label{eq:Liang_F}
\end{equation}
where $C^{0}$ here is an average of the quiet-Sun cross-covariance function. They computed the analytic signal of $F$, $s[F] = F+ \mathrm{i} {\mathcal H}[F]$, where $\mathcal H[F]$ denotes the Hilbert transform of $F$. Finally, the amplitude $\aL13$ was given by the maximum of the envelope of the analytic signal 
\begin{equation}
\aL13 =  \max_t \; |s[F](t)| \; .
\label{eq:aL13_adef}
\end{equation}

It is natural to ask if there are any relations between this definition and our linear one. As we are measuring a perturbation to the amplitude, let us define 
a perturbed amplitude for \cite{2013A&A...558A.129L} as
\begin{equation}
\delta \aL13 = \aL13 - 1.
\end{equation}
In Appendix~\ref{sec:app_L13}, we show that 
\begin{equation}
\delta \aL13 \geq \delta A, \label{eq:comp_aL13_alin}
\end{equation}
and that we have equality if the maximum of $F$ is obtained at $t = 0$ and  the cross-covariances are symmetric with respect to time. Thus, if the time lag between the quiet-Sun and the measured cross-covariance is small both definitions are equivalent, otherwise our linear definition will always give a smaller amplitude than the definition from \cite{2013A&A...558A.129L}.

\subsection{Noise properties}
\label{sec:noise}

Using the quiet-Sun noise data from  
\cite{2015A&A...581A..67L} based on the solar noise model (realization noise) 
from \cite{2004ApJ...614..472G}, we calculate the amplitudes in a quiet-Sun noise field. For this calculation, we use only the realization 
noise field with the p$_1$-ridge filter, and center-to-annulus cross-covariance functions for a 10-Mm-radius annulus. Figure \ref{fig:cc_example_p1_noise} shows the reference cross-covariance function (i.e., averaged cross-covariance over the field of view), the window function and the weight functions, as well as one example cross-covariance function with its amplitude and travel-time perturbations calculated using our definition.

\begin{figure}[]
\centering
{\includegraphics[]{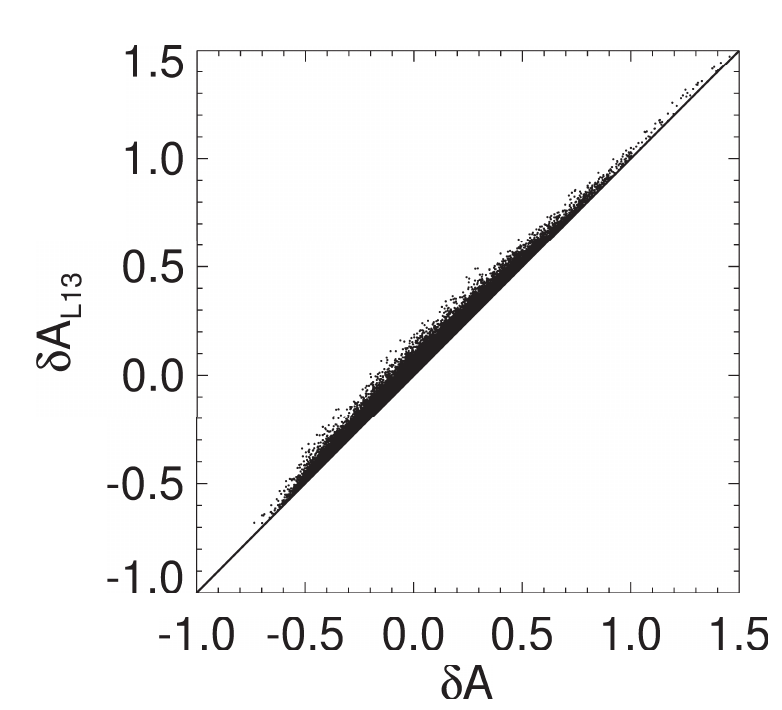}}
\caption{Scatter plot between the linear definition of the amplitude $\delta A$ and $\delta\aL13$ for the case of a (quiet-Sun) realization noise calculation.
These two amplitude measurements show a good correlation:
Pearson's correlation coefficient is 0.996.
This plot also shows $\delta \aL13 \ge \delta A$, which is expected from Eq.~(\ref{eq:comp_aL13_alin}).}
\label{fig:scatplot_p1_noise_a}
\end{figure}

Figure \ref{fig:scatplot_p1_noise_a} compares
$\delta A$ and $\delta\aL13$.
In this plot the two amplitude measurements show a good correlation; Pearson's correlation coefficient is 0.996 in this case.
Thus, both definitions seem to have a similar behavior with noise. However, an advantage of the linear definition is that the noise can be estimated easily. Due to linearity, we can use the noise covariance matrix from \cite{2004ApJ...614..472G} and \cite{2014A&A...567A.137F} by replacing the weight function for travel-time $W_\tau$ by the weight function for amplitude $W_A$. In particular, the measurement of the amplitude is unbiased and its standard deviation decreases as $1/\sqrt{T}$  as the observation time $T$ increases.  
We also note that Fig.~\ref{fig:scatplot_p1_noise_a} shows that $\delta\aL13$ is greater than $\delta A$ which is consistent with Eq.~(\ref{eq:comp_aL13_alin}).

\begin{figure}
\centering
{\includegraphics[]{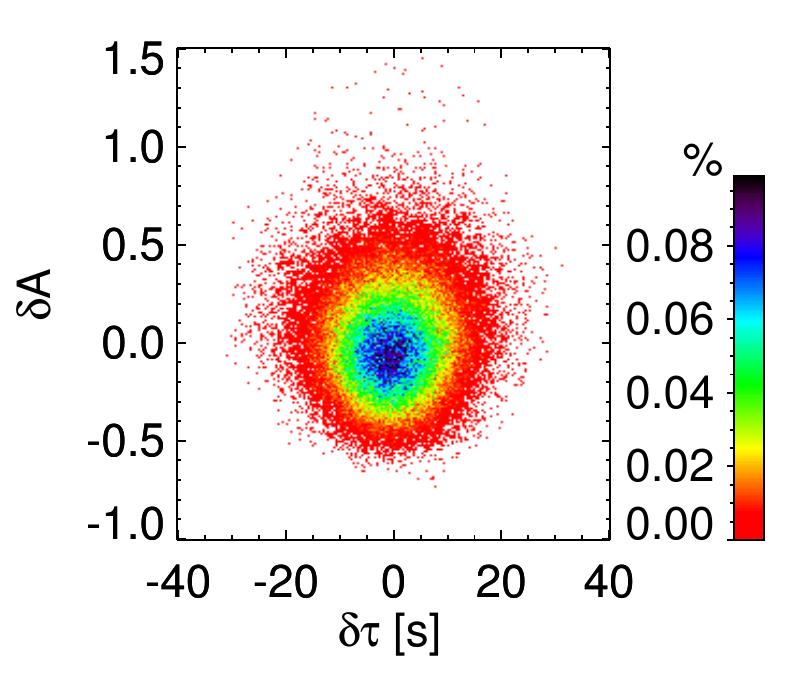}}
\caption{Scatter plot of the amplitudes  $\delta A$ and travel times $\delta \tau$ for the case of the (quiet-Sun) realization noise calculation.  
The color indicates the percentage of points in each grid cell of size 0.4 s for $\delta \tau$ and 0.01 for $\delta A$.  The total number of the data points is 51465. }
\label{fig:scatplot_p1_noise_at}
\end{figure}

An important question is the independence of the travel-time and amplitude measurements. As a first step, we look at the correlation between the noise for travel-time and amplitude for the case of realization noise. 
Figure~\ref{fig:scatplot_p1_noise_at} shows the scatter plot between $\delta\tau$ and $\delta A$. As expected the mean value is zero for both  $\delta\tau$ and $\delta A$, confirming that the linear definitions are unbiased. Moreover, the distribution of the points indicates that the amplitude and travel time noise measurements are 
largely uncorrelated; 
the Pearson's correlation coefficient between $\delta\tau$ and $\delta A$ is $0.02$. 

\section{Example: cross-covariance amplitude around a
sunspot}\label{sec:example}

In this section, we use our new definition to compute the amplitude perturbations from the cross-covariance function around a sunspot measured by \cite{2013A&A...558A.129L} and show that the results are consistent with their measurements.

\subsection{Observations used by \cite{2013A&A...558A.129L}}

We use the cross-covariance functions calculated by \cite{2013A&A...558A.129L} .
These cross-covariance functions were calculated from a 9-day Dopplergram dataset around
active region NOAA 9787 obtained by MDI \citep{1995SoPh..162..129S}.
Here we will use the same coordinate system as \cite{2013A&A...558A.129L}; the sunspot center is the origin of the local coordinates, $(x,y)=(0,0)$.
Since they calculated cross-covariance functions in a transverse cylindrical equidistant projection centered at the sunspot center (i.e., the origin), 
the wave front, if in the quiet Sun, would be parallel to the $y$-axis (great circle on the Sun) in these  $x$--$y$ coordinates.

As for the cross-covariance function definition, 
\cite{2013A&A...558A.129L} replaced Eq.~(\ref{eq:CCdef}) by
\begin{eqnarray}
C_n(x,y,t) = \int_{-T/2}^{T/2} \langle \phi_n(x=x_0, t^\prime) \rangle \; \phi_n (x,y,t^\prime+t)  \;  \mathrm{d}t^\prime  , \label{eq:def_Cn}
\end{eqnarray}
where $(x,y)$ is the spatial position on the surface, $t$ is the time lag, $\phi_n$ is the filtered Doppler velocity, $\langle \phi_n \rangle$ is the average of $\phi_n$ over $y$ along the line $x=x_0 \equiv -43.73 \ \mathrm{Mm}$, and with $T=1 \ \mathrm{day}$. The Doppler velocity signals, $\phi_n$, are obtained by applying ridge filters to select the modes with radial order  $n=0, 1, 2, 3, 4$.
The cross-covariance function is then averaged over nine days and over 
rotations around the sunspot center (i.e., the origin) as well. 
The details of the ridge filters and other parameters are found in \cite{2013A&A...558A.129L}.

In their calculation they defined the quiet-Sun cross-covariance $C^{0}_n (x,t)$ for each mode with radial order $n$ at each $x$  as the average of $C_n(x,y,t)$
over the line segments where $100 \ \mathrm{Mm} <|y|<200 \ \mathrm{Mm}$ and also $\sqrt{x^2+y^2 } < 262 \  \mathrm{Mm}$ (to avoid the effect of the apodization). We use this $C^{0}_n$ as the reference
cross-covariance for our linear travel-time and amplitude calculations as well.

\subsection{Comparison of amplitudes measured using the two methods}\label{sec:measurements}

\begin{figure*}
\resizebox{\hsize}{!}
{\includegraphics{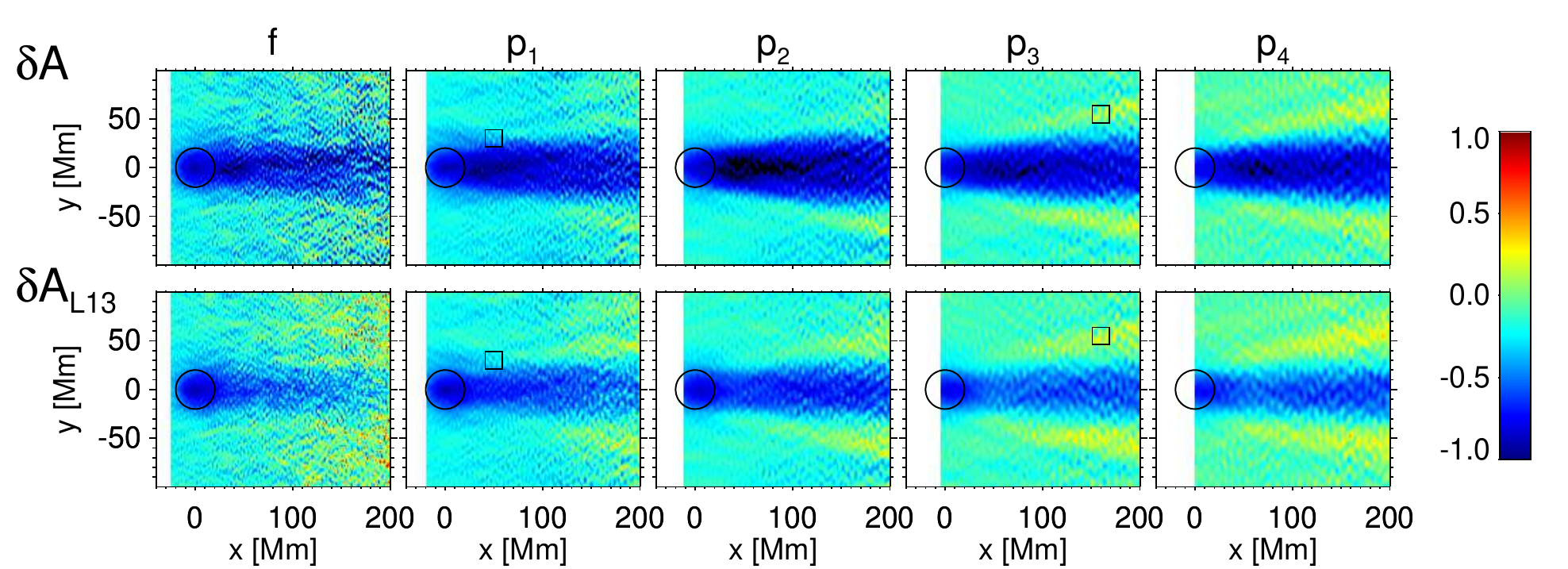}}
\caption{Observed amplitude perturbations around the sunspot NOAA 9787 with respect to the quiet Sun using SOHO/MDI observation.
From left to right the results for f, p$_1$, p$_2$, p$_3$, and p$_4$ modes are shown. 
The black circle shows the outer boundary of the penumbra.
The upper panels show the linear amplitude, $\delta A$,
while the lower panels are replots of Fig.~2 of \cite{2013A&A...558A.129L}  (for better comparison, here we show $\delta \aL13 = A_{\rm L13}-1$ instead of $\aL13$ as in Fig.~2 of \cite{2013A&A...558A.129L} ).
The central points of the squares in p$_1$ and p$_3$ panels are the positions of the cross-covariance functions shown in Fig.~\ref{fig:ZhiChao_cc}.
The measurement results
in the near field are not displayed in the plots;
the near field is defined as $x-x_0 < 3\lambda$, where $\lambda$ is the typical wavelength for each mode, and $x_0 = -43.73 \ \mathrm{Mm}$ is the leftmost point of 
these plots over which $\langle \phi_n \rangle$ in Eq.~(\ref{eq:def_Cn}) is averaged (See Table 1 in \cite{2013A&A...558A.129L} for details). 
}
\label{fig:ZhiChao_refSPy12_apdmap}
\end{figure*}

\begin{figure}
{\includegraphics[width=\linewidth]{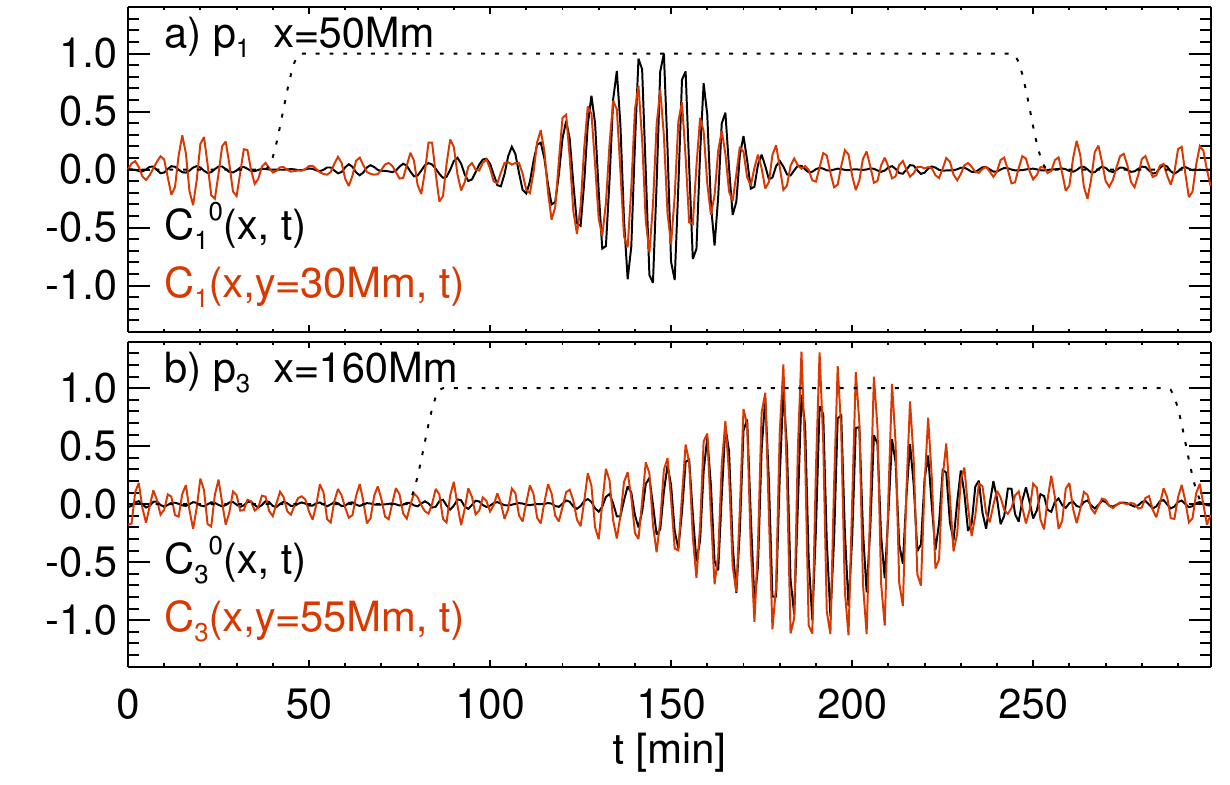}}
\caption{Example cross-covariance functions used in \cite{2013A&A...558A.129L}. Panels a) and b) are p$_1$ cross-covariances at $(x,y)=(50 \ \mathrm{Mm},30 \ \mathrm{Mm})$ and p$_3$ cross-covariances at $(x,y)=(160 \ \mathrm{Mm}, 55 \ \mathrm{Mm})$, respectively
(see the squares drawn in Fig.~\ref{fig:ZhiChao_refSPy12_apdmap}). In each panel $C_n^{0}(x,t)$ is drawn in a black solid line, and $C_n(x,y,t)$ is shown in a red line and normalized by the maximum of the absolute value of $C_n^{0}(x,t)$.
The dotted lines are the window functions.
The amplitudes for the cross-covariance in panel a) are $\delta A =-0.45$ and $\delta \aL13=-0.28$,
while for panel b)  $\delta A =0.19$ and $\delta \aL13=0.25$.
}
\label{fig:ZhiChao_cc}
\end{figure}

Figure \ref{fig:ZhiChao_refSPy12_apdmap} shows  
two amplitude measurements ($\delta A$ and $\delta \aL13$) 
around the sunspot.  
In both amplitude maps,
the amplitude behind the sunspot shows a deficit along the $x$-axis and small enhancement off-axis on each side. 
These are mainly explained by geometrical spreading.
As shown by \cite{2013A&A...558A.129L}, using 2D ray tracing,
the refraction of the waves resulting from the increased sound
speed in the sunspot is responsible for most of the observed amplitude variations, namely the amplitude reduction along the $y=0$ axis (defocusing) and the amplitude enhancement away from the axis (wavefront folding). See Fig.~3.7-5 of \cite{2003.Stein.Wysession} and Chapter~15.4 of \cite{1998DahlenTromp} as well as \cite{2000JGR...10519043N} 
for geophysical analogs.
In addition, part of the amplitude perturbations might be due to wave absorption in the sunspot as a result of the mode conversion of incoming p modes into downward propagating magneto-acoustic wave in the strong magnetic field \citep[e.g.,][]{1992ApJ...391L.109S, 1997ApJ...486L..67C, 2008SoPh..251..291C}.

In Fig.~\ref{fig:ZhiChao_refSPy12_apdmap} 
the linear ($\delta A$) and non-linear ($\delta \aL13$) amplitude measurements show reasonable agreement,
even in the rather faint details away from the $x$ axis. 
Fig.~\ref{fig:ZhiChao_cc} shows a few examples of local changes in the cross-covariance function. 
The change in the amplitude is negative in Fig.~\ref{fig:ZhiChao_cc}a ($\delta A =-0.45$ and $\delta \aL13=-0.28$) where the amplitude of the perturbed cross-covariance is smaller than the quiet-Sun cross-covariance,
and positive ($\delta A = 0.19$ and $\delta \aL13 = 0.25$) in Fig.~\ref{fig:ZhiChao_cc}b. We note that $\delta A$ is not the ratio between the maximum of the amplitudes of $C$ and $C^0$ but is the value that gives the best fit in the least-square sense between $C$ and $C^0$ in the whole window function.

\begin{figure}
{\includegraphics{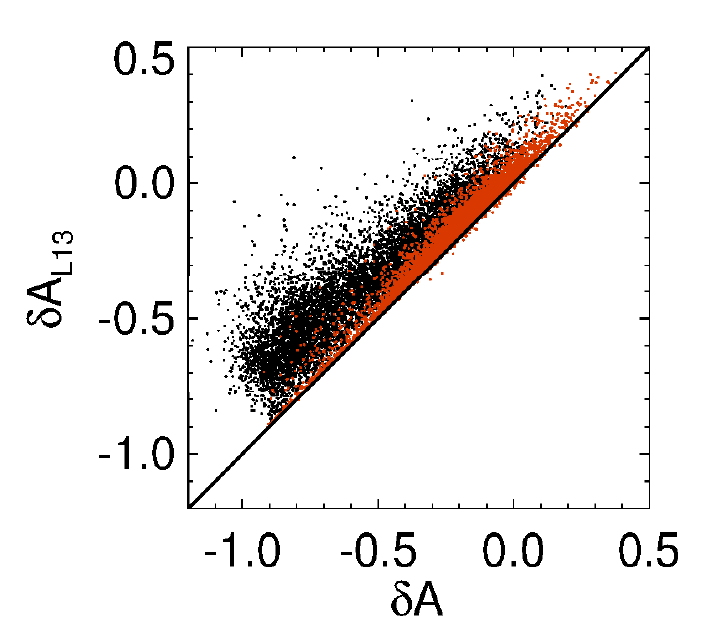}}
\caption{Scatter plot of amplitudes for the p$_1$ 
cross-covariance functions. The data points in the field of the view of the p$_1$ panels of Fig.~\ref{fig:ZhiChao_refSPy12_apdmap} are used for this plot; the Pearson's correlation coefficient is $0.91$.
The red points are the points
in the area where $|\delta \tau_{\mathrm{L13}} |< 20 \ \mathrm{s}$; The Pearson's correlation coefficient for these red data points is $0.98$.}
\label{fig:scatplot_p1_a}
\end{figure}

To better compare both measurements, Fig.~\ref{fig:scatplot_p1_a} shows the scatter plot between $\delta A$ and $\delta\aL13$. The correlation is very good, in particular when the absolute value of the travel-time perturbation in \cite{2013A&A...558A.129L}'s result, $|\delta \tau_{\mathrm{L13}}|$, is less than 20~s (red points in  Fig.~\ref{fig:scatplot_p1_a}). In this case, Pearson's correlation coefficient between $\delta A$ and $\delta \aL13$  is above $0.95$ not only for p$_1$ (shown in Fig.~\ref{fig:scatplot_p1_a}) but also for all other modes. 
As expected from Eq.~(\ref{eq:comp_aL13_alin}), the values of $\delta\aL13$ are always greater than the linear measurement, $\delta A$. We note moreover that there is no saturation of the linear measurement, and the linear and nonlinear values are consistent even for strong amplitude variations, although the measurements where the perturbation is larger show larger deviation from the line with a slope of one.

\section{Conclusions}\label{sec:conclusions}

We have developed a convenient linear procedure to measure the amplitude
of the cross-covariance function of solar oscillations.  We have shown that this new definition qualitatively reproduces the amplitude measurements computed by the more complex and non-linear definition of \citet{2013A&A...558A.129L} even for strong perturbations in a sunspot.

The main advantage of our definition of the amplitude given by Eq.~(\ref{eq:amp_def}) is the linear relation between cross-covariance and amplitude. It allows easy generalization of the procedure developed in \cite{2002ApJ...571..966G} for computing travel-time sensitivity kernels under the Born approximation to compute kernels for amplitudes as well.  The linear definition presented here also enables straightforward calculation of the amplitude noise covariance matrix by following the approach \cite{2004ApJ...614..472G} and \cite{2014A&A...567A.137F}.  Thus, all the tools are in place for linear forward modeling and inversions using amplitude measurements in addition to, or instead of, travel times.

The conclusions from Earth seismology show that amplitudes contain information that is not in the travel times alone.  In particular, the kernel calculation by \cite{2005SeismicEarth_37} shows that the sensitivity kernels for amplitude and travel time have different spatial sensitivity. 

For problems where linear theory is insufficient, numerical simulations of wave propagation are an important tool for interpreting helioseismic measurements \citep[e.g.,][]{2008SoPh..251..291C}. Typically, these simulations have been used to predict the travel times that would be expected for a particular model of physical conditions in the solar interior (e.g., a sunspot) and then these modeled travel times are compared with travel times determined from observations.  Comparison of forward-modeled amplitudes with observed amplitudes will, in general, provide additional constraints on models.  \cite{2013A&A...558A.130S} used wave propagation simulations to compute the sensitivity of travel times to small changes in a model sunspot; the extension of this work to amplitudes will help quantify the diagnostic power of amplitudes in the context of sunspot seismology.  

Another case where amplitudes may be an important diagnostic is in characterizing small-scale turbulence.  
Scattering from small-scale (granulation scale) turbulence is, in some regimes, thought to be an important contributor to wave attenuation
\citep{1998ApJ...505L..55D} 
and thus should play a role in determining amplitudes.

\begin{acknowledgements}
We thank Zhi-Chao Liang and Thomas Duvall for useful discussions.
The German Data Center for SDO, funded by the German Aerospace Center (DLR), provided the IT infrastructure for this work.
K.N. and L.G. acknowledge support
from EU FP7 Collaborative Project “Exploitation of Space Data for Innovative Helio- and Asteroseismology” (SpaceInn). 
SOHO is a project of international cooperation between ESA and NASA.
\end{acknowledgements}

\bibliographystyle{aa} 

\begin{appendix}

\section{Derivation of the amplitude of the noise-free cross-covariance function} \label{sec:ap_2parader}

Here we briefly summarize the derivation of the two-parameter fitting formula and discuss when the two measurements (the amplitude and the travel time) are decoupled.

Let us look for $a=A$ and $t=\tau$ that minimize $X(a,t)$ (Eq.~(\ref{eq:2para_chisq})): 
\begin{eqnarray}
\frac{\partial X}{\partial a}(A,\tau) = 0 \ \text{ and } \ \frac{\partial X}{\partial t}(A,\tau) = 0.
\end{eqnarray}
Let us write
\begin{eqnarray}
\overline{C}(t) = \C0(t)+\delta \overline{C}(t) \ , \
a = a^0 + \delta  a \  , \  t = t^0 +\delta t,
\end{eqnarray} 
where the superscript $0$ is for the reference medium and $\delta$ for the perturbation. The zeroth-order cost function is given by
\begin{equation}
X^0(a^0, t^0) = \int^{+\infty}_{-\infty} w(t^\prime) \left[ \C0(t^\prime)  - a^0 \C0
(t^\prime - t^0)  \right]^2  \mathrm{d}t^\prime 
\end{equation}
and is minimal for
\begin{equation}
a^0 = A^0 = 1 \quad \textrm{and} \quad t^0 = \tau^0 = 0.
\end{equation}
Then, $\delta A$ and $\delta \tau$ are obtained by solving:
\begin{eqnarray}
\frac{\partial X}{\partial a}(1+\delta A,\delta\tau) = 0 \ \text{ and } \ \frac{\partial X}{\partial t}(1+\delta A,\delta\tau) = 0.
\end{eqnarray}
To first order, we obtain
\begin{eqnarray}
  \left(    \begin{array}{cc}
      I_1 & -I_2  \\
      I_2 & I_3  
    \end{array}  \right)
\left(   \begin{array}{c}
      \delta  A \\
      \delta \tau  
    \end{array} \right)
&= &
\left(   \begin{array}{c}
      \int^{+\infty}_{-\infty} w(t) \C0(t) \delta \overline{C}(t)   \mathrm{d}t \\\
      \int^{+\infty}_{-\infty} w(t)  \bigdot{C}{}^0(t) \delta \overline{C}(t) \mathrm{d}t \ 
    \end{array} \right),
    \label{eq:2para_matrix}
 \end{eqnarray} 
where 
\begin{eqnarray}
I_1 &=& \int^{+\infty}_{-\infty} w(t) \left[\C0(t)\right]^2 \mathrm{d}t \ , \\
I_2 &=& \int^{+\infty}_{-\infty} w(t) \C0(t) \bigdot{C}{}^0 (t)  \mathrm{d}t \ , \label{eq:2para_I2} \\
I_3 &=& -\int^{+\infty}_{-\infty} w(t) \left[\bigdot{C}{}^0 (t)\right]^2  \mathrm{d}t \ .
\end{eqnarray}

Let us show that the term $I_2$ can be dropped 
if the window function is chosen such that it isolates a wavepacket. We can write
\begin{eqnarray}
I_2 &=&  \int^{+\infty}_{-\infty}  w(t) \C0 (t) \bigdot{C}{}^0 (t)\mathrm{d}t = \frac{1}{2}\int^{+\infty}_{-\infty} w(t)  \frac{\mathrm{d}}{  \mathrm{d}t}\left[\C0(t)\right]^2 \mathrm{d}t \  \nonumber \\
&=&\frac{1}{2}  \left[w(t)\left[\C0(t) \right]^2 \right] ^{+\infty}_{-\infty} -\frac{1}{2} \int^{+\infty}_{-\infty} \bigdot{w}(t)\left[{\C0} (t)\right]^2 \mathrm{d}t \ .
\end{eqnarray}
The first term is zero, 
since $\C0(t)$ tends to zero at infinity.  
The second term is small if $w(t)$ is constant or varies slowly over the duration of the wavepacket, for example, by using a smooth rectangular function as shown in Fig.~\ref{fig:toyPlot}.

Therefore, we define $\delta A$ and $\delta \tau$ as
\begin{eqnarray}
\delta  A &=& \int^{+\infty}_{-\infty}W_{A}(t) \delta \overline{C}(t)  \mathrm{d}t   \label{eq:2para_da_WC}\\
\delta \tau &=& \int^{+\infty}_{-\infty} W_{\tau}(t) \delta \overline{C}(t) \mathrm{d}t \label{eq:2para_dtau_WC}
\end{eqnarray} 
with the weight functions given by Eqs.~(\ref{eq:F2para_a_weight_new}) and (\ref{eq:F2para_t_weight_new}).

The weight functions $W_{A}$ and $W_{\tau}$  are 
identical to what we would obtain from two independent single-parameter fits. In other words, 
the weight function $W_{\tau}$ is identical 
to the weight function obtained by minimizing 
$X(t)$ defined by Eq.~(\ref{eq:GB1para_chisq}).
This is exactly the same as in \cite{2002ApJ...571..966G}. 
In the same manner, the weight function $W_{A}$
is identical to the weight function obtained by minimizing $X(a)=\int^{+\infty}_{-\infty} w(t) \left[\overline{C}(t) - a \C0 (t) \right]^2 \mathrm{d}t $.

\section{Derivation of the amplitude in the noisy case}
\label{sec:ap_GB04_2para}

Here we extend the method used in \cite{2004ApJ...614..472G} in the case of noisy data. This is important
since the observations of finite duration can be so noisy that the minimization problem defined by Eq.~(\ref{eq:2para_chisq}) does not make sense.
We show that this leads to our linearized formulae (Eq.~(\ref{eq:2para_matrix})).

In this case we introduce a smoothed cross-covariance function, $C_{\epsilon}$, defined by
\begin{equation}
C_{\epsilon} (t) = \epsilon C(t) + (1-\epsilon) \C0
(t),
\end{equation}
where $\epsilon$ is a small positive number. In the limit $\epsilon \rightarrow 0^+$ (no noise), $ C_{\epsilon} \rightarrow \C0$.

We are looking for $a_\epsilon = A_\epsilon$ and $t_\epsilon = \tau_\epsilon$ that minimize the cost function 
\begin{equation}
X_{\epsilon}(a_\epsilon, t_\epsilon) = \int^{+\infty}_{-\infty}  w(t) \left[ C_{\epsilon} (t)  - a_\epsilon \C0
(t - t_\epsilon)  \right]^2  \, \mathrm{d}t .
\end{equation}
Let us write
\begin{equation}
t_\epsilon = t_\epsilon^0 + \epsilon \delta t_\epsilon \quad \textrm{and} \quad a_\epsilon = a_\epsilon^0 + \epsilon \delta a_\epsilon, 
\end{equation}
where $t_\epsilon^0$ and $a_\epsilon^0$ represent the zeroth-order terms in $\epsilon$ while $\delta t_\epsilon$ and $\delta t_\epsilon$ are of order 1.
At zeroth order, the cost function is given by
\begin{equation}
X_{\epsilon}^0(a_\epsilon^0, t_\epsilon^0) = \int^{+\infty}_{-\infty} w(t) \left[ \C0(t)  - a_\epsilon^0 \C0
(t - t_\epsilon^0)  \right]^2  \, \mathrm{d}t 
\end{equation}
and is thus minimal for
\begin{equation}
a_\epsilon^0 =A_\epsilon^0 = 1 \quad \textrm{and} \quad t_\epsilon^0 = \tau_\epsilon^0 = 0 \ .
\end{equation}
In order to find $\delta A_\epsilon$ and $\delta \tau_\epsilon$, let us minimize the cost function $X_\epsilon$ 
\begin{equation}
 \frac{\partial X_{\epsilon}}{\partial a_\epsilon } ( A_{\epsilon}, \tau_{\epsilon})= 0  \; \text{ and } 
 \frac{\partial X_{\epsilon}}{\partial t_\epsilon} (A_{\epsilon}, \tau_{\epsilon}) = 0 \ .
\end{equation}
Keeping only the first-order terms in $\epsilon$, we obtain
\begin{eqnarray}
\int_{-\infty}^\infty  w  \left( C - \C0 - \delta A_\epsilon \C0 + \delta \tau_\epsilon \bigdot{C}{}^0\right)  \C0 \, \mathrm{d}t &=& 0 , \\
\int_{-\infty}^\infty  w  \left( C - \C0 - \delta A_\epsilon \C0 + \delta \tau_\epsilon \bigdot{C}{}^0 \right) \bigdot{C}{}^0 \, \mathrm{d}t &=&  0 .
\end{eqnarray}
At the limit $\epsilon \rightarrow 0^+$, it leads to the noise-free formulation given by Eq.~(\ref{eq:2para_matrix}) and thus $\delta \tau_\epsilon \rightarrow \delta \tau$ and $\delta A_\epsilon \rightarrow \delta A$. So the formulation given by Eq.~(\ref{eq:2para_matrix}) makes sense even in the case of noisy observations.

\section{Comparison of amplitudes $A$ and $\aL13$} \label{sec:app_L13}

In this appendix, we compare our linear definition with the one proposed by \cite{2013A&A...558A.129L} where the amplitude was obtained as the maximum of the analytic signal $s[F]$ of the function $F$ defined by Eq.~(\ref{eq:Liang_F}), namely
\begin{equation}
\aL13  = \max_t |s[F](t)| = \max_t |F(t) + \textrm{i} \mathcal{H}[F](t)|,
\end{equation}
where  $\mathcal{H}[F]$ denotes the Hilbert transform of $F$.
As the maximum of an expression over all $t$ is larger than the value at $t=0$, we have
\begin{equation}
\aL13 \geq |F(0) + \textrm{i} \mathcal{H}[F](0)| \geq F(0). \label{eq:majoration}
\end{equation}
Adding and subtracting $\C0$ in the definition of the function $F$, we obtain
\begin{eqnarray}
F(0) &=& \frac{\int^{+\infty}_{-\infty}  w(t^\prime) \left(C(t^\prime) + \C0(t^\prime) - \C0(t^\prime) \right) C^0(t^\prime) \mathrm{d}t^\prime }{\int^{+\infty}_{-\infty}   w(t^\prime) [C^0(t^\prime)]^2 \mathrm{d}t^\prime} \nonumber \\
&=& 1 + \int^{+\infty}_{-\infty} W_A(t^\prime) \delta C(t^\prime)  \mathrm{d}t^\prime \nonumber \\
&=& 1 + \delta A. \label{eq:F0}
\end{eqnarray}
Then, combining Eqs.~(\ref{eq:majoration}) and (\ref{eq:F0}), we obtain
\begin{equation}
\aL13 \geq A \ ,
\end{equation}
and thus Eq.~(\ref{eq:comp_aL13_alin}).

Moreover if the maximum of $F$ is obtained at $t = 0$, then
$\aL13 = |F(0) + \textrm{i} \mathcal{H}[F](0)|$.
If additionally, $C$ and $\C0$ are symmetric then $F$ is symmetric, and $\mathcal{H}[F]$ is antisymmetric which implies that $\mathcal{H}[F](0) = 0$. In this case, $\aL13 = F(0) = A$, and our linear definition coincides with the one from \cite{2013A&A...558A.129L}.

\end{appendix}

\end{document}